\documentclass[preprint]{aastex}
\usepackage{psfig}
\received{} 
\accepted{} 
\journalid{}{} 
\articleid{}{} 
\shortauthors{Mitchell, R.C. et~al.}
\shorttitle{$^{56}$Ni Mixing in SN 1987A}

\def\ifundefined#1{\expandafter\ifx\csname#1\endcsname\relax}

\newif\ifpdf
\ifx\pdfoutput\undefined
   \pdffalse      
\else
   \pdfoutput=1   
   \pdftrue
\fi

\def\la{\mathrel{\hbox{\rlap{\hbox{\lower4pt\hbox{$\sim$}}}\hbox{$<$}}}}
\def\ga{\mathrel{\hbox{\rlap{\hbox{\lower4pt\hbox{$\sim$}}}\hbox{$>$}}}}

\newcommand{\be}{\begin{eqnarray}}
\newcommand{\ee}{\end{eqnarray}}

\ifundefined{ensuremath}\def\ensuremath#1{\relax\ifmmode{#1}}
\else${#1}$\fi\else\relax\fi
\ifundefined{nuc}\def\nuc#1#2{\relax\ifmmode{}^{#1}{\protect\text{#2}}
\else${}^{#1}$#2\fi}\else\relax\fi

\newcommand{\etal}{et al.}

\newcommand{\kmps}{km~s$^{-1}$}

\newcommand{\msol}{\ensuremath{{\textrm{M}_\odot}}}

\newcommand{\xni}{\ensuremath{\textrm{X}_{\textrm{Ni}}}}
\def\ang{\hbox{\AA}}

\def\tstd{\ensuremath{\tau_{\textrm{std}}}}

\newcommand{\nick}{$^{56}$Ni}
\newcommand{\phoe}{{\tt PHOENIX}}

\begin{document}

\title{\nick\ Mixing in the Outer Layers of SN 1987A}

\author{Robert C.~Mitchell, E.~Baron, David Branch}

\affil{Department of Physics and Astronomy, University of Oklahoma, 440 West
Brooks, Norman, OK 73019-0261, USA}
\email{mitchell@mail.nhn.ou.edu, baron@mail.nhn.ou.edu, branch@mail.nhn.ou.edu}

\author{Peter Lundqvist}

\affil{Stockholm Observatory, SE--133~36 Saltsj\"obaden,
Sweden}

\email{peter@astro.su.se}

\author{Sergei Blinnikov}

\affil{Institute of Theoretical and Experimental Physics, 117218,
Moscow, Russia}
\email{blinn@sai.msu.su}

\author{Peter H.~Hauschildt}

\affil{Department of Physics and Astronomy \& Center for Simulational Physics,
University of Georgia, Athens, GA 30602, USA}
\email{yeti@hal.physast.uga.edu}

\author{and}

\author{Chun S.~J.~Pun}

\affil{Laboratory for Astronomy and Solar Physics,
NASA/GSFC, Code~681, Greenbelt, MD~20771}
\email{pun@congee.gsfc.nasa.gov}

\begin{abstract}
     Supernova 1987A remains the most well-observed and well-studied
supernova to date. Observations produced excellent broad-band photometric and
spectroscopic coverage over a wide wavelength range at all
epochs. Here, we focus on the very early spectroscopic observations.
Only recently have numerical models
been of sufficient detail to accurately explain the
observed spectra.  In SN~1987A, good agreement has been found between
observed and synthetic spectra for day one, but by day four, the predicted
Balmer lines become much weaker than the observed lines.  We
present the results of work based on a radiation-hydrodynamic model by
Blinnikov and collaborators.  Synthetic
non-LTE spectra generated from this model
by the general radiation transfer code \phoe\ strongly support the
theory that significant mixing of \nick\ into the outer envelope is
required to maintain strong Balmer lines.
Preliminary results suggest a lower limit to the average nickel mass of
$1.0 \times 10^{-5}$
solar masses is required above 5000 \kmps\ by day four.  \phoe\ models
thus have the potential
to be a sensitive probe for nickel mixing in the outer layers of a supernova.
\end{abstract}

\section{Introduction\label{intro}}

Until the explosion of Supernova 1987A in the LMC, computer models of
supernovae were usually calculated with assumptions that allowed only 
the prediction of bolometric light curves.  The matter temperature was usually
assumed to be equal to the radiation temperature, and grey opacities
were used \citep{sn87arev89,blinn87a99a}.
Often, the models did not
account for the effects of radiation on the hydrodynamic gas flow, or
the effects of sphericity on the spectrum \citep{schm90}.

Observations of SN~1987A produced extensive
broad-band photometry, especially during its earliest phase of
expansion.  This motivated the development of models to predict its spectral
and photometric evolution. In our analysis here we make use of the
optical data obtained at CTIO \citep{ctio87a88} and the extensive IUE
data which was obtained, re-reduced and analyzed by \citet{punetal95}.

While SN~1987A led to the growth of observational data, 
so too did it lead to an improvement in
theoretical models.
\citet{hflh87} modeled the early phase of SN~1987A, using a pure
hydrogen atmosphere and accounting for non-LTE (Local Thermodynamic
Equilibrium).  \citet{eastkir89} modeled the first ten days of the
SN~1987A explosion. H~I and He~I were modeled in NLTE, while metal
lines were treated as pure scattering LTE lines. \citet{tak91} used a
pure H/He envelope in NLTE and attempted to understand the lineshapes
of the Balmer lines.

Our work here utilized a hydrodynamic model developed by Blinnikov and
collaborators using the {\tt STELLA} software package
\citep{blinn93j98,blinn87a99a,blinn87a00}.  This model calculated the
light curve for up to six months after the explosion, and includes
allowance for time-dependent, multi-group radiation hydrodynamics,
monochromatic scattering effects, and the effects of spectral lines on
the opacity \citep{blinn87a00}. The hydrodynamical models were based
on the stellar evolution models of \citet*{saio88b}, \citet*{saio88a},
and \citet{nhpr88}.  The procedure for synthesizing SN~1987A spectra
from Blinnikov \etal's model and the results are described in
\S~\ref{models}.  Here we concentrate on the fact that, by day 4.52,
significant discrepancies begin to appear in the synthetic spectra of
SN~1987A with respect to the observed spectra.  In \S~\ref{day4} we
review earlier works which noted this discrepancy.
In \S~\ref{nickel} we show that nickel mixing at early times seems to
be the only way to make the synthetic spectra agree with observations.
In a subsequent paper, we will discuss NLTE effects with regard to the
results of \citet{blinn87a00}.

\section{Models\label{models}}

\subsection{\phoe}

\phoe\ is a general radiation transfer code that computes temperature,
opacity, and level populations for each of typically 50--100 radial
zones in a moving 
stellar envelope.
\phoe\ allows the user to calculate level populations for a multitude of
different nuclear species in LTE or NLTE \citep{hbjcam99,short99}.
For this study, the 
following species were calculated in NLTE: H~I, He~I, He~II, O~I, Ne~I, Na~I,
Mg~II, Si~II, S~II, Ca~II, Fe~I, Fe~II, and Fe~III.

The Blinnikov \etal\ hydrodynamic model consists of an ejecta envelope
divided into 300 zones, with compositions defined for each layer
\citep{blinn87a99a,blinn87a00}.  The outer layers of this model were re-zoned
onto the 75-zone grid that is used by a modified version of \phoe, preserving
the compositions in velocity-space.  The inner boundary condition was taken
to be diffusive, which is well justified in these calculations.
The gamma-ray energy deposition rate, which is the energy
in gamma-rays deposited in the matter in the envelope per unit mass per
unit time, is calculated for each new zone.  The new grid was chosen so as to
maintain both composition and density profiles with adequate
resolution.

\subsection{Spectral Synthesis}

\begin{figure}
\begin{center}
\leavevmode
\plotone{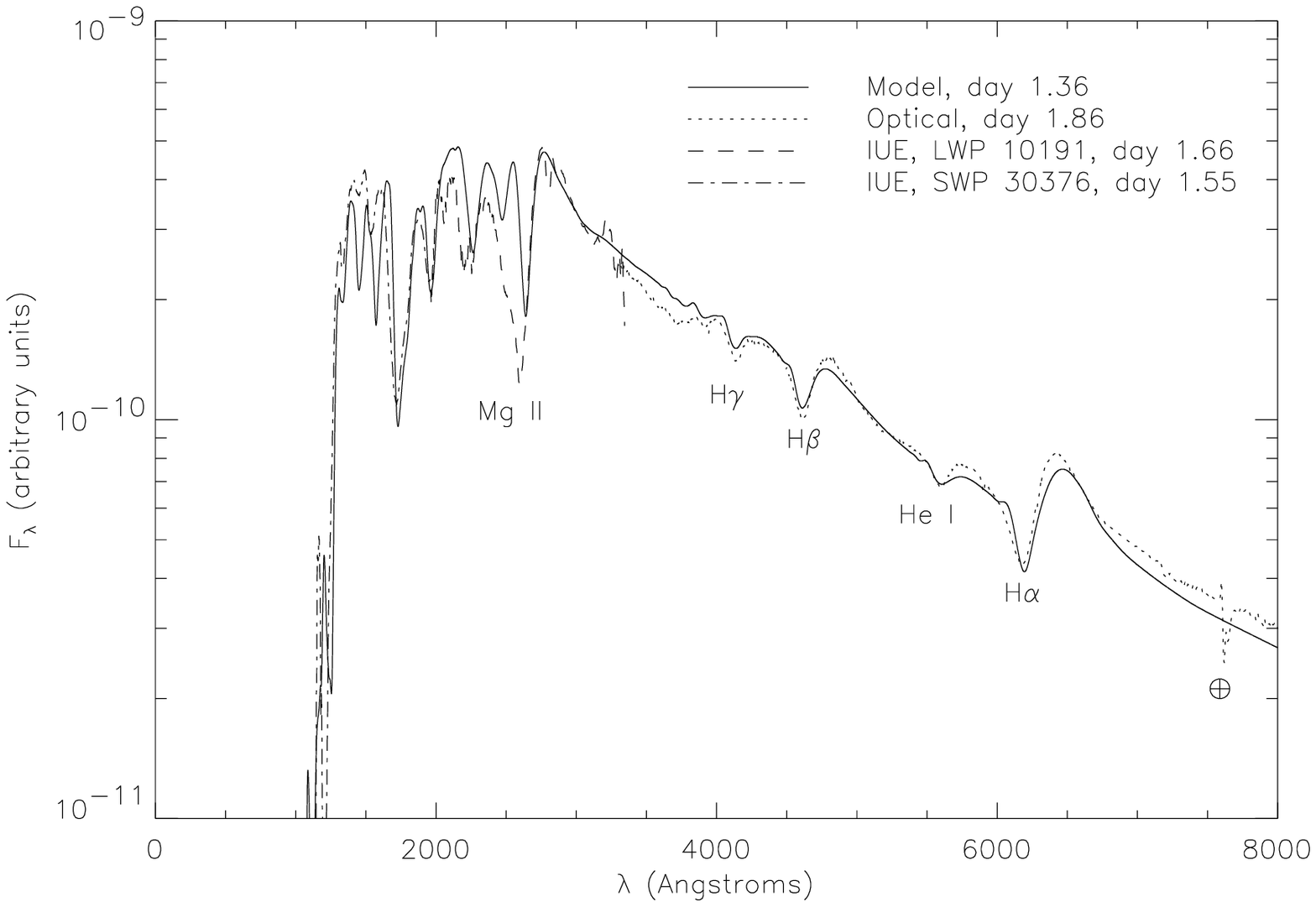}
\caption{\label{fig:1opt} \phoe\ model spectrum for day~1.36, plotted
against observed optical and UV spectra.  Important optical lines include
H$\alpha$ through H$\delta$, and He~I~$\lambda$5876.  All optical
spectra are taken from the CTIO archive \citep{ctio87a88}.  All UV
spectra are from IUE \citep{punetal95}.}
\end{center}
\end{figure}

\begin{figure}
\begin{center}
\leavevmode
\plotone{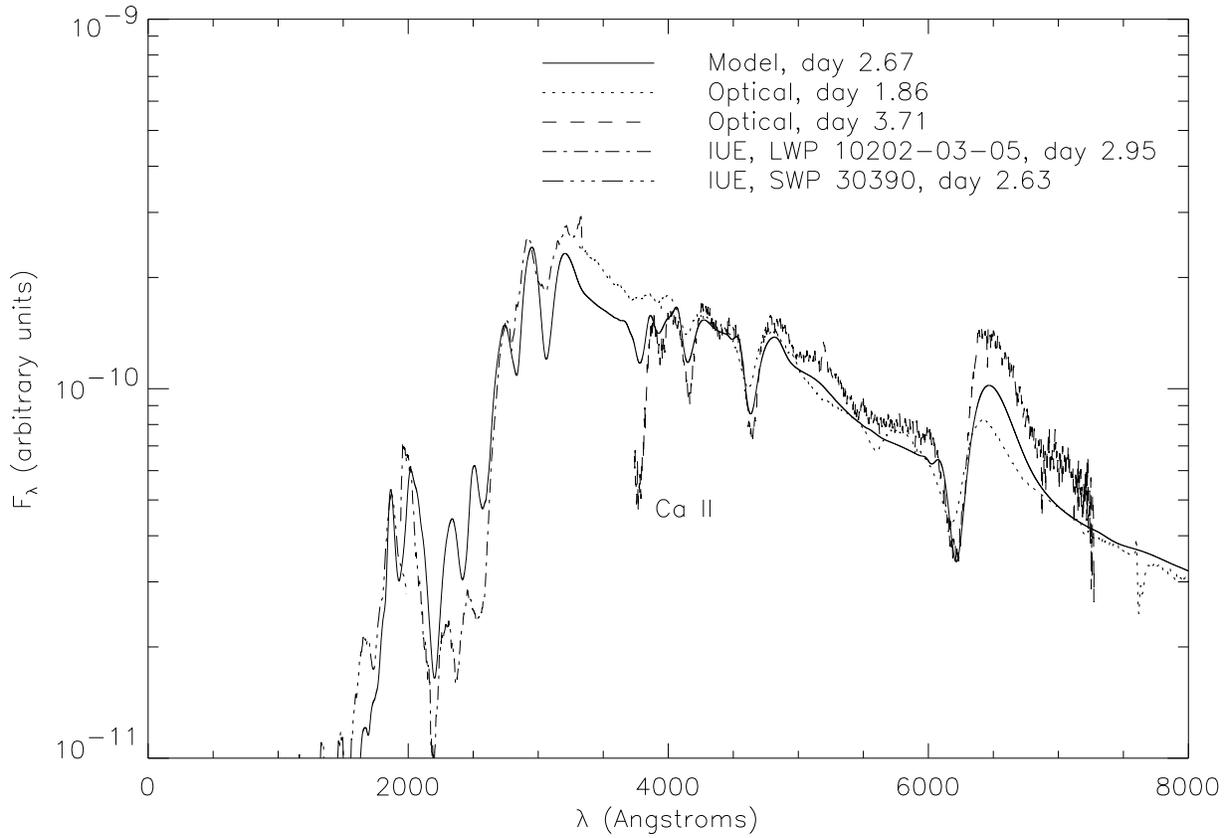}
\caption{\label{fig:2opt} \phoe\ model spectrum for day~2.67.  Note that
He~I~$\lambda$5876 has virtually disappeared while Ca~II~H\&K lines are
beginning
to emerge.  No observed optical spectrum for day~2.67 (Feb 26) was available
\citep{ctio87a88}.}
\end{center}
\end{figure}

\begin{figure}
\begin{center}
\leavevmode
\psfig{file=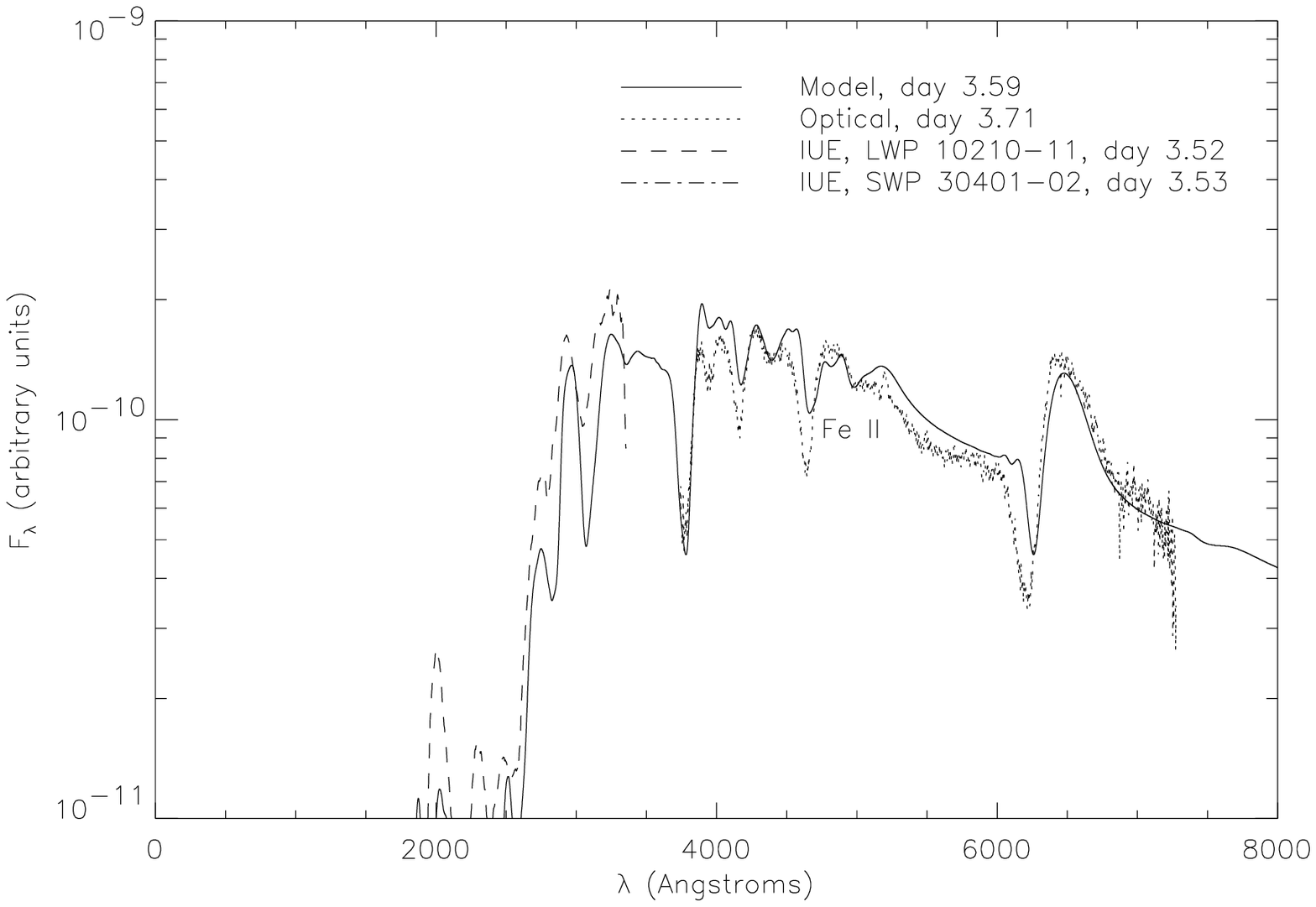,width=14cm,height=9cm}
\psfig{file=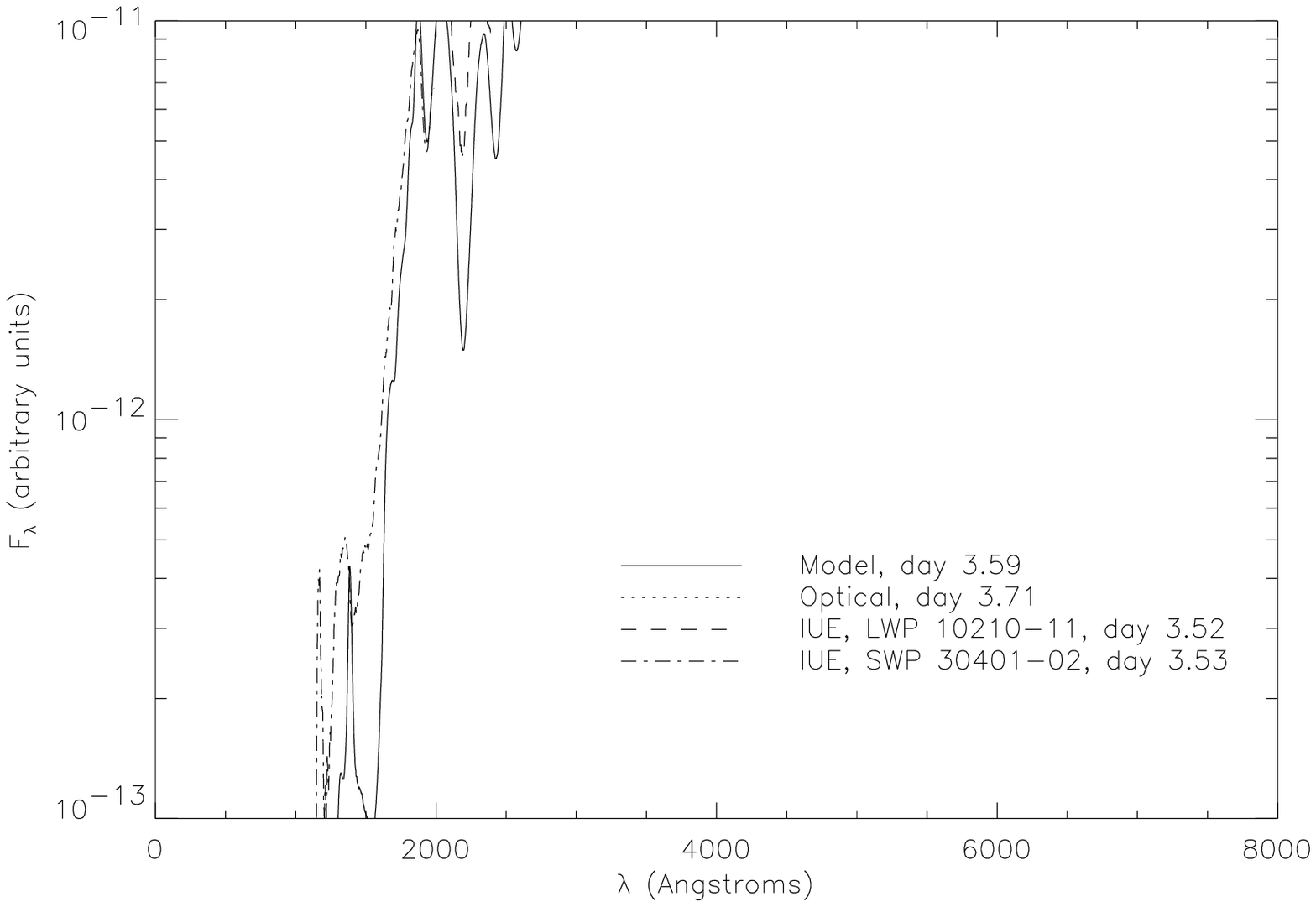,width=14cm,height=9cm}
\caption{\label{fig:3opt} \phoe\ model spectrum for day~3.59.  The Ca~II lines
have become stronger, and there is now much greater line blanketing in the
UV.  Note that the predicted H$\alpha$ line is now noticably weaker than
the observed line.}
\end{center}
\end{figure}

\begin{figure}
\begin{center}
\leavevmode 
\psfig{file=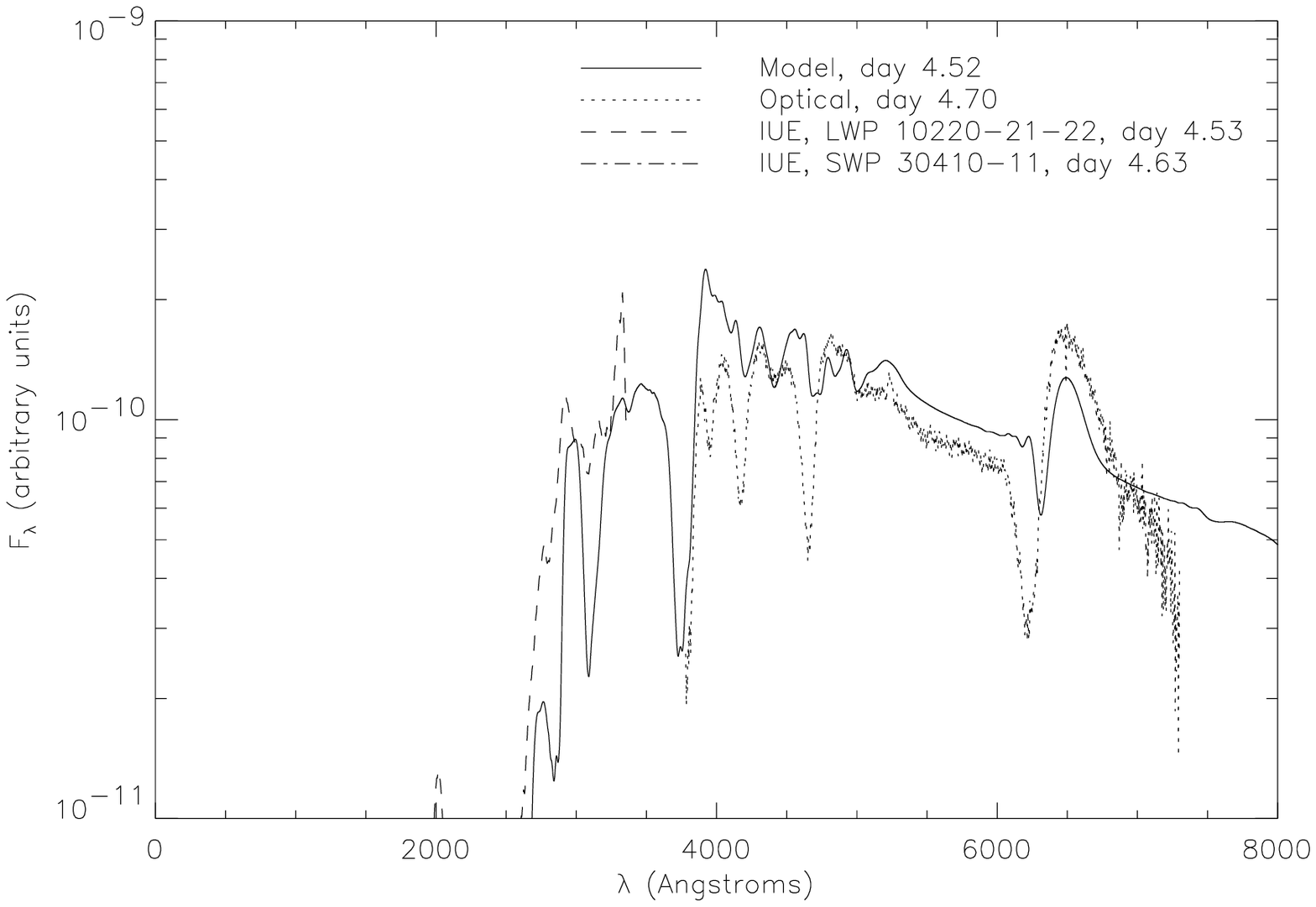,width=14cm,height=9cm}
\psfig{file=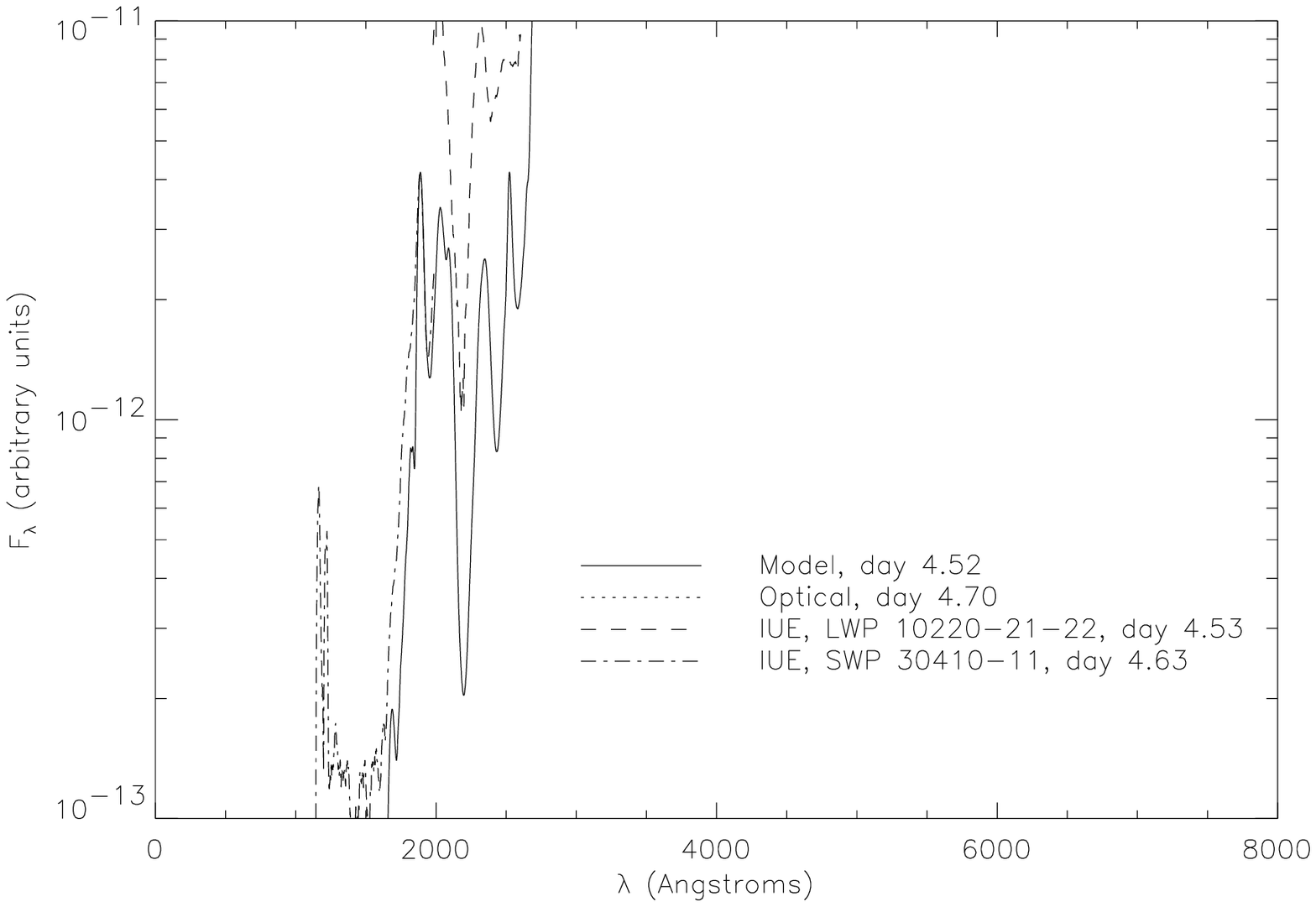,width=14cm,height=9cm}
\caption{\label{fig:4opt} \phoe\ model spectrum for day~4.52.  All synthetic
Balmer lines are now much weaker than the observed lines and are slightly
redshifted.  The observed lines are bluer since they formed at a higher
velocity.}
\end{center}
\end{figure}

Figures~\ref{fig:1opt}--\ref{fig:3opt} show
the spectra resulting from \phoe\ computations on days 1.36, 2.67, and 3.59,
respectively, of SN~1987A.  The fits between the IUE spectra and the synthetic
UV spectra are very good, an indicator of the importance of NLTE effects in
the supernova envelope.  [\citet{blinn87a00} noted that the predicted UV flux
5--10 days after shock breakout was much greater than the actual flux when
computed in LTE.]  For days 1.36
and 2.67, the agreement between the
synthetic and observed optical spectra are also reasonably good.  However, by
day 3.59 a discrepancy in the strengths of the Balmer lines begins to appear
in the synthetic spectra, most notably H$\alpha$, even when NLTE effects are
included.
Figure~\ref{fig:4opt} shows the \phoe-generated optical/near-UV spectrum
for day 4.52 compared with the observations \citep{ctio87a88,punetal95}, and
here the Balmer lines are clearly
much weaker in the synthetic spectrum than they are in the observed spectrum.

\section{Weak Lines\label{day4}}

\citet{schm90}, using various hydrodynamic models, noted the hydrogen
atoms appeared to be sufficiently
excited to create strong lines, despite the fact that the photospheric
temperature had dropped below the ionization threshold ($T \approx 6000 K$).
They could offer no
explanation for this discrepancy.  The pure H/He non-LTE model of
\citet{tak91} also resulted in weak Balmer lines, as did a model by
\citet{phhens94}, which used an earlier version of \phoe\ with H~I,
He~I, Mg~II, 
and Ca~II in NLTE and metal line blanketing in LTE.  These two groups
suggested that differences between the theoretical and actual density
distributions might be responsible.

\begin{figure}
\begin{center}
\leavevmode
\plotone{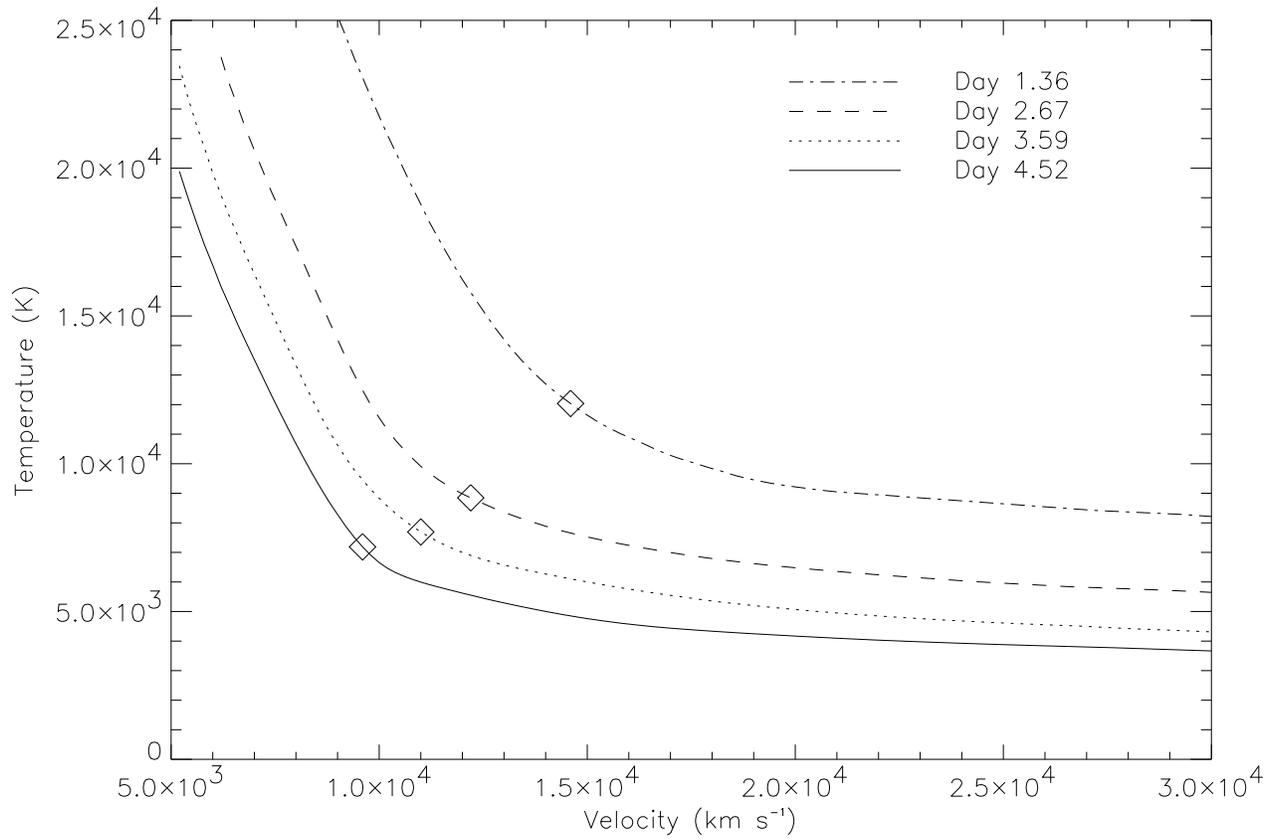}
\caption{\label{fig:5vel} Temperature structure vs. velocity for
days~1.36 through 4.52.  The diamonds denote the approximate locations
of $\tstd=1$ for each model.}
\end{center}
\end{figure}

Figure~\ref{fig:5vel} shows temperature vs.\ velocity for
Blinnikov \etal's models at days 1.36 to 4.52.  The diamonds on the graph show
where \tstd\ is approximately equal to 1 in each model
($\tstd\equiv$ total continuum optical depth at 5000~\ang).
Assuming the line-forming
region lies in the vicinity of $\tstd=1$, note that for days 1 and 2,
the gas temperature is well above the ionization threshold for
hydrogen, thus ensuring the presence of strong Balmer lines.  (A good rule of
thumb is that one sees the strongest lines from the species that is just
below the dominant ionization stage.  For example, Si~II lines are
strongest if 
Si~III is the dominant stage, etc.)  By day
3, the temperature is beginning to approach the threshold at
$\tstd=1$, and at day 4 the temperature is such that sufficient
ionization can no longer be assumed in the line-forming region.

\begin{figure}
\begin{center}
\leavevmode
\psfig{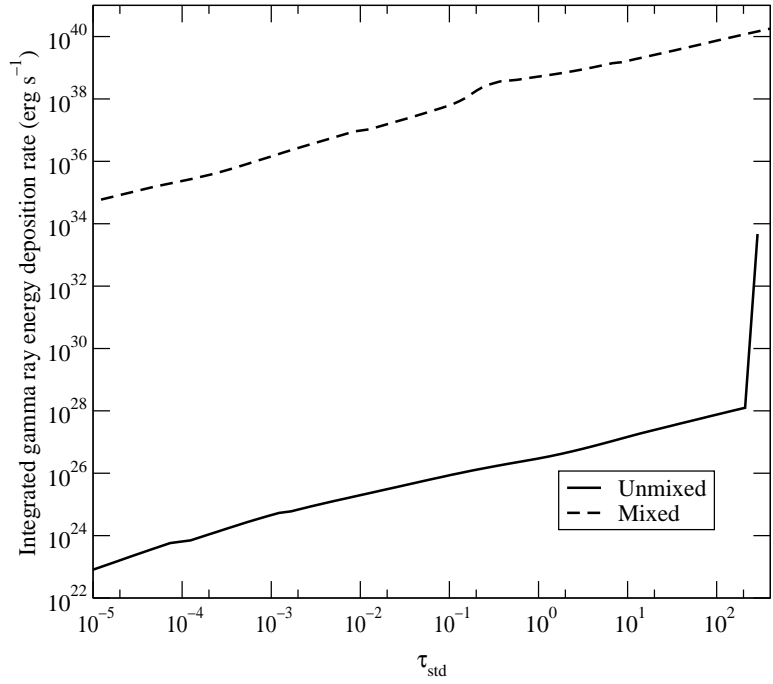}
\caption{\label{fig:6dep} The total gamma-ray energy deposition rate
vs. $\tstd$ for day~4.52.  The quantity plotted is the integrated
energy deposited in the envelope starting at the outer edge. The
unmixed deposition rate is obtained from Blinnikov \etal's model
\citep{blinn87a99a,blinn87a00} and the mixed rate is that obtained
with our assumed mixing.}

\end{center}
\end{figure}

Figure~\ref{fig:6dep}  shows the gamma-ray energy deposition
 predicted by Blinnikov \etal's hydrodynamic model. The total energy
 deposited per second is shown as a function of $\tstd$ and the
 integral is performed from the outside proceeding inward. Until we
 reach the 
innermost layers of the envelope model ($\tstd > 100$), the deposition is
negligible, and therefore hydrogen atoms cannot become sufficiently excited
through the absorption of gamma rays.

\section{Nickel Mixing\label{nickel}}

\subsection{Model Results}

In order for the synthetic Balmer lines to match the strength of the observed
lines,  either higher matter temperatures (to thermally
populate the Balmer states) or higher radiation temperatures (to
radiatively populate the Balmer states in NLTE) would be
required. Both of these would 
significantly degrade the quality of the fit, since they would both
lead to model spectra that would be significantly bluer than the
observed spectrum. A third way to strengthen the Balmer lines is to
populate the states non-thermally, via fast electrons created from the
deposition of gamma-rays produced by the radioactive decay of \nick.
In our models this is controlled by the gamma-ray energy deposition rate.
In order to obtain a larger energy deposition rate, more mixing of
radioactive \nick\ into the
supernova's outer envelope than exists in
Blinnikov \etal's model \citep{blinn87a99a,blinn87a00} is required.
In \phoe\, this can be 
accomplished by replacing the self-consistent energy deposition rate from
the Blinnikov \etal\ model with an ad hoc function that simulates a gamma-ray
population created by excess nickel mixing.  To simulate mixing, we assume a
uniform mass fraction of \nick\ distributed homogeneously (i.e.,
${\xni} = $~constant), and local deposition
of gamma-rays.
The new energy deposition rate is therefore much larger (and smoother) than
the original (Figure~\ref{fig:6dep}), allowing for more hydrogen ionization
and excitation in the line-forming region.
Comparisons between the synthetic spectrum and the actual spectrum provide
a sensitive probe for how much \nick\ must be mixed into the supernova
envelope.

\begin{figure}
\begin{center}
\leavevmode
\psfig{file=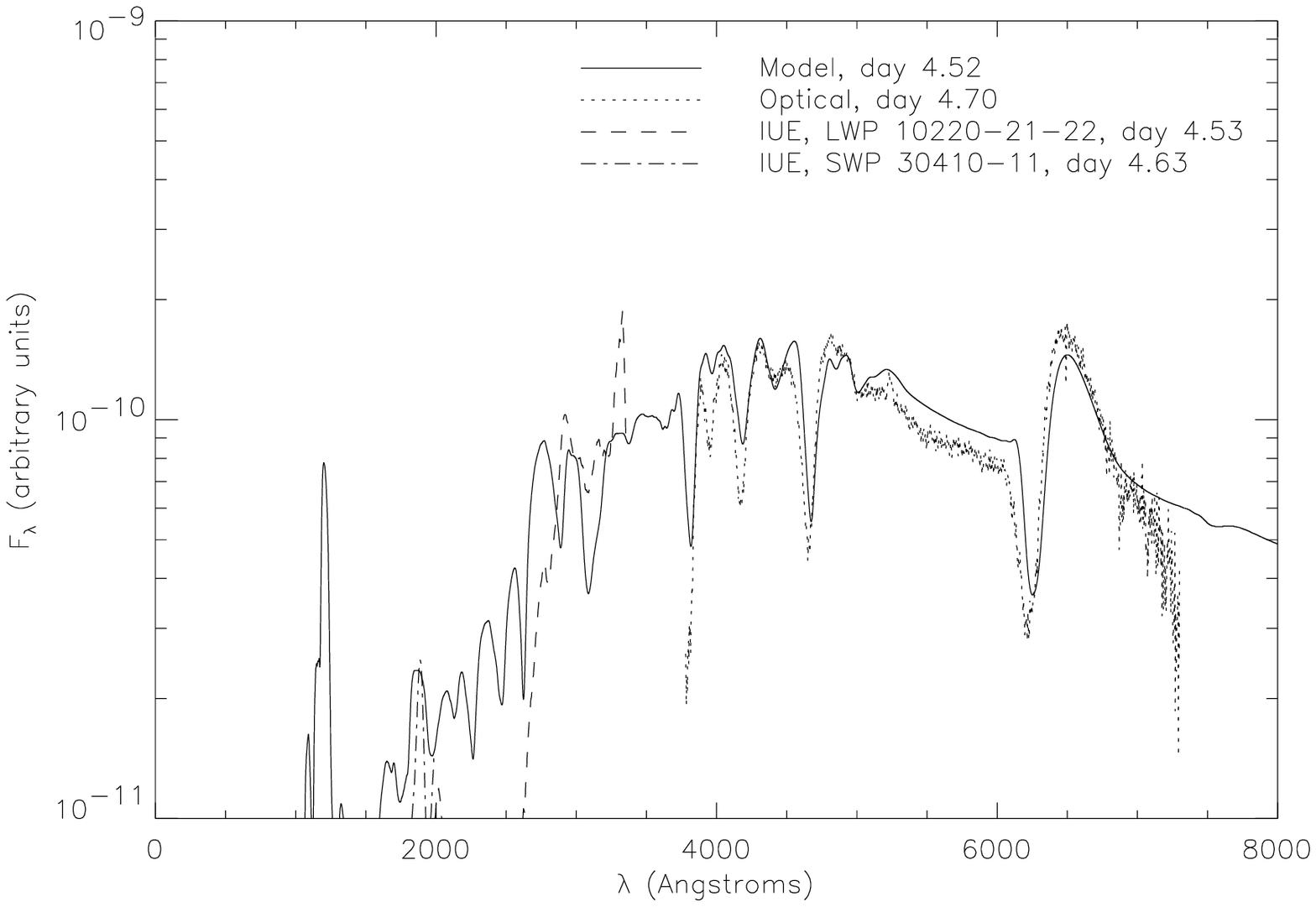,width=14cm,height=9cm}
\psfig{file=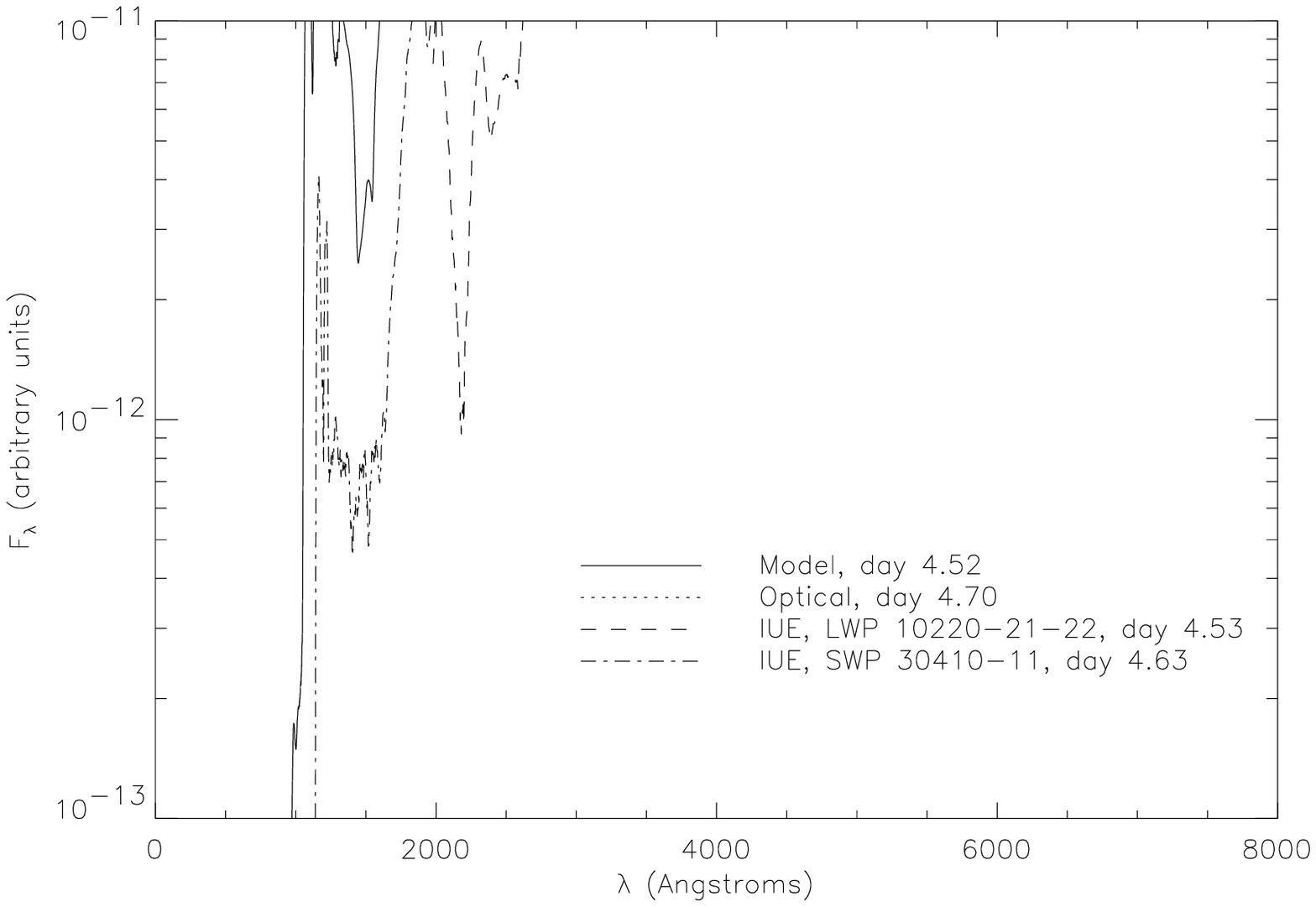,width=14cm,height=9cm}
\caption{\label{fig:7nopt} \phoe\ model spectrum for day~4.52, with gamma-ray
deposition calculated assuming local deposition due to a 
constant nickel mass fraction of $1.0 \times 10^{-3}$ everywhere in
the envelope.} 
\end{center}
\end{figure}

Figure~\ref{fig:7nopt} shows the results of modeling the day-4.52
spectrum with constant gamma-ray deposition, resulting from a nickel
mass fraction of $\xni = 1.0 \times 10^{-3}$ in the envelope.
The fit between the 
predicted lines and the actual lines is much better, illustrating the
validity of this model.  A mass fraction of ${\xni} = 1.0 \times
10^{-3}$ distributed thoughout our model atmosphere with a mass of about
$10^{-2}$~\msol\ 
corresponds to a requirement of approximately $(1.0-1.2) \times 10^{-5}$
\msol\ of nickel to create enough gamma rays above the line
forming region of the envelope.  The mass is low (and corresponds to a
lower boundary in velocity space of about 5000~\kmps), because at
these early times the density is high and the continuum optical depth
is $> 100$ already at this mass/velocity point, so that spectrum
formation occurs in the very outer 
layers of the ejecta. At later times we see deeper due to 
geometrical dilution. Although the model produced by \phoe\
utilized a constant nickel mass fraction (so that the number density
of nickel follows the rather steep density gradient), this by no means
precludes the 
possibility of inhomogeneous mixing. In the case of that the nickel is
in clumps, our nickel mass is a lower limit (since the clumps could
self-absorb gamma rays) and our results indicate
that at least some of the clumps must be optically thin to the gamma
rays that are produced.

In preliminary lower resolution studies, we found that the synthetic
H$\alpha$ line in day 4.52 was slightly redshifted with respect
to the actual H$\alpha$.  This was also observed in synthetic spectra for days
3.71 and 5.76 created by the radiative transfer code {\tt SYNOW}
\citep{jb90}.  This indicates that when the line strengths are correct, so
is their velocity and demonstrates the diagnostic power of detailed
spectral synthesis.  

\subsection{More Evidence for Nickel Mixing}

The mechanism for the mixing of \nick\ into the envelope is the
subject of intense debate, and won't be settled until the
core-collapse supernova mechanism is better understood.
\citet{kpjm00} and \citet{kpm01} suggested that nickel mixing into the
hydrogen shell would be suppressed in Type II supernovae due to strong
deceleration at the He/H boundary.  Nevertheless, evidence exists
to support the results of the \phoe\ models.  Early detection of X-rays
\citep{dotanietal87} and gamma-rays
\citep{matzetal88}, the expansion velocities in infrared
line widths \citep{ericketal88,witte89} and in $^{56}$Co gamma-ray lines
\citep{barthetal89}, the detection of H velocities as low as 800
km s$^{-1}$ at day 221 \citep{hflich88}, and the late detection of the
He~I~10830 line \citep{FM87A99} all suggest that significant mixing has
occurred in the ejecta [see \citet{sn87arev89}, \citet{sn87arev93} and
references therein].

Two-dimensional, axially symmetric models calculated by
\citet{fma91} and by \citet{hb91,hb92}
suggested that \nick\ could be mixed into the envelope via finger-like
structures that are produced shortly after the explosion by instabilities
in the matter behind the shock front.  By $t = 90$~days, these fingers of
nickel had broken up and expanded through radioactive heating into
low-density bubbles of cobalt or iron, the decay products of \nick.

Inhomogeneous mixing has already been suggested as an
explanation for the so-called ``Bochum event,'' an emission satellite on the
red side of the H$\alpha$ line in days 20--100.  \citet{utrobin87a95}
suggested 
that a high-velocity clump of \nick\ caused an asymmetric over-excitation
of hydrogen in the outer envelope.  Recent studies by  Chandra 
have detected the presence of diffuse high-velocity Fe bubbles at the outer
edge of the Cassiopeia A supernova remnant.  \citet{hughescasa00} noted
the detection of Fe bubbles at the outer edge of the supernova ejecta,
outside the Si-rich layers, and suggested that their diffuse nature could be
explained by radioactive heating from decaying nickel.

Models of the composition of SN~1987A in the nebular phase by
\citet{kozfran98a}, based on the results of various explosion models,
used a number fraction of iron of $2 \times 10^{-5}$ in the hydrogen layer,
with a filling factor $f \approx 0.17-0.70$.  This corresponds to a
mass fraction on the order of $10^{-3}$.  The results of the
simulations of \citet{kozfran98b} indicate that the iron mass fraction
should be within a factor of 2 of the predicted value.
\citet{lietal93} suggest that the light curve of SN~1987A during the
nebular phase was dominated by $\approx 0.078$ \msol\ of $^{56}$Co in
the form of 50 to 100 clumps.  Their models contained nickel clumps
that were initially opaque to gamma rays with $f < 0.01$, accounting
for $\approx 1\%$ of the total envelope mass.  Gamma-ray absorption
heated the clumps, causing them to expand until $f = 0.5$.  The nickel
mass fraction and total mass values predicted by the \phoe\ runs are
well within the limits imposed by these two studies.

Detection of the He~I~10830 line at times after 10 days suggests that
helium is also being re-ionized by \nick\ decay.  \citet{FM87A99}
applied spectral synthesis models to SN~1987A and noted that the best fit
to the helium line required at least 3\% of the total \nick\ mass
(0.07 \msol) be mixed to above 3000 \kmps, but that \nick\
concentrations should be negligible above $\approx 4000$ \kmps.
\citet{fassiaetal98} carried out a similar analysis of the
Type~II supernova 1995V, where they noted the
best fit to the He line required the \nick\ density to be constant up
to 650--1150 \kmps, then fall off by a power law of 8--9 (consistent
with our mixing assumption).  Only
$10^{-6}$ \msol\ of \nick\ was required above the helium photosphere
($\approx 4250$ \kmps\ at day 69) in the SN~1995V models.  \citet{burvr95},
using Monte Carlo calculations of gamma-ray transport in clumpy debris
models of SN~1987A's envelope, suggest that up to 50\% of the total \nick\
mass in SN~1987A should remain below 1000 \kmps.

\citet{hbhfc94} suggest that significant iron/nickel
inversion may be the result of neutrino-driven convection within the
fraction of a second just prior to the supernova explosion.
Convection would destroy the spherical
symmetry of the elemental distribution, possibly creating the iron/nickel
clumps predicted and/or detected in later stages.  This theory is supported
by 2D hydrodynamic simulations performed by \citet{kpm01}, although in their
models the nickel was strongly decelerated at the He/H interface.

\section{Further Research\label{further}}

Future work with the Blinnikov \etal\ 1987A model will involve using
\phoe\ to synthesize the entire time series of the supernova up to
around day 100.  The detailed time series will indicate the location
of nickel in the entire envelope, since the spectra probe deeper as
time goes on due to the geometrical thinning of the envelope.
\phoe\
spectra for this period may shed light on 
the later-time phenomena discussed in \S~\ref{nickel}.  Once
the whole time series of SN~1987A is modeled, the synthetic spectra
will be applied to the Spectral-fitting Expanding Atmosphere Method
(SEAM) in order to calculate the distance to SN~1987A, and hence to
the LMC \citep{b93j1,b93j2,b93j3,b93j4}. Our calculations will also be
useful to ascertain whether a 1-D representation of SN~1987A is
sufficient. So far both breakout models \citep{blinn87a00} and our
preliminary calculations here indicate this to be the case.

\acknowledgments We thank Claes Fransson for support and helpful
discussions. This work was supported
in part by NSF grants  AST-9731450, NASA
grant NAG5-3505, and an IBM SUR grant to the University of Oklahoma;
and by NSF grant AST-9720704,
NASA ATP grant NAG 5-8425 and LTSA grant NAG 5-3619 to the University
of Georgia. SB is supported in Russia by the grant RFBR
99-02-16205. PHH was supported in part by the P\^ole 
Scientifique de Mod\'elisation Num\'erique at ENS-Lyon. Some of the
calculations presented here were 
performed at the  San Diego
Supercomputer Center (SDSC), supported by the NSF, and at the National
Energy Research Supercomputer Center (NERSC), supported by the
U.S. DOE.  We thank both these institutions for a generous allocation
of computer time.


\clearpage

\end{document}